\documentclass[preprint,aps]{revtex4}

\usepackage{graphicx}
\usepackage{dcolumn}

\begin{document}

\preprint{Lebed-PRL-2008}

\title{Paramagnetic Intrinsic Meissner Effect in 
Layered Superconductors}

\author{A.G. Lebed$^*$}

\affiliation{Department of Physics, 
University of Arizona,
1118 E. 4-th Street, Tucson, AZ 
85721, USA}

\begin{abstract}
Free energy of a layered superconductor with $\xi_{\perp} < d$
is calculated in a parallel magnetic field by means of the Gor'kov 
equations, where $\xi_{\perp}$ is a coherence length perpendicular 
to the layers and $d$ is an inter-layer 
distance.
The free energy is shown to differ from that in the textbook Lawrence-Doniach 
model at high fields, where the Meissner currents are found to create 
an unexpected positive magnetic moment due to shrinking of the Cooper 
pairs "sizes" by a magnetic field.
This paramagnetic intrinsic Meissner effect in a bulk is suggested 
to detect by measuring in-plane torque, the upper critical field, 
and magnetization in layered 
organic and high-T$_c$ superconductors as well as in 
superconducting
superlattices.  
\end{abstract}

\maketitle
  
\pagebreak

The Meissner diamagnetic effect is known to be the most important 
property of a superconducting phase and is responsible for 
destruction of superconductivity both in type-I and type-II 
superconductors [1].
Meanwhile, as shown by us in Refs. [2-4] and independently 
by Tesanovic, Rasolt, and Xing in Ref. [5], quantum effects of an 
electron motion in a magnetic field result in the appearance of 
a qualitatively different phenomenon - superconductivity surviving 
in high magnetic fields in layered [2-4] and isotropic three-dimensional 
[5] type-II superconductors.
In particular, it was shown [2-4,6] that in a layered conductor in 
a parallel magnetic field, where the Landau levels quantization 
is impossible, some other quantum effects - the Bragg reflections - 
play an important role.
These quantum effects result in a "two-dimensionalization" (i.e., 
$3D \rightarrow 2D$ dimensional crossover) of an open electron 
spectrum in an arbitrary weak parallel magnetic field.
This is known [6,7] to cause the field-induced spin-density-wave (FISDW) and 
field-induced charge-density-wave (FICDW) instabilities 
in layered quasi-one-dimensional (Q1D) 
conductors.
More complicated $3D \rightarrow 1D \rightarrow 2D$ dimensional
crossovers are shown [8,9] to be responsible for the experimentally 
observed non-trivial angular magnetic oscillations in a metallic phase
of different layered organic conductors, including Interference Commensurate 
[8] and Lebed Magic Angle 
[9]  oscillations.

As shown in Refs.[2-4,7], the similar $3D \rightarrow 2D$ dimensional 
crossovers have to be responsible for a stabilization of a superconducting 
phase in layered Q1D [2,4] and quasi-two-dimensional (Q2D) [3] 
conductors since 2D superconductivity is not destroyed  in a parallel
magnetic 
field.
More precisely, it is shown [2-4] that: (i) the quantum effects make the 
upper critical field to be divergent, $H^{\parallel}_{c2}(T) \rightarrow \infty$ 
as $T \rightarrow 0$, and (ii) there is some critical filed, $H^*$, 
above which superconducting temperature grows in an increasing 
magnetic
field. 
Such superconducting phase with $d T_c / dH >0$ is called the 
Reentrant Superconductivity (RS) [2-5].
The original predictions [2-4] have been theoretically confirmed by a 
number of studies [10-14], including a study [14], which takes into account
a possibility  of a pure 2D phase 
transition. 
Despite of a great success of $3D \rightarrow 1D \rightarrow 2D$
and $3D \rightarrow 2D$ dimensional crossovers concepts in the 
explanations of magnetic properties in a metallic [7-9], the FISDW [6,7], 
and the FICDW [7] phases of organic conductors, so far there has been no  
evidence that superconducting temperature can grow in high magnetic 
fields due to the the quantum $3D \rightarrow 2D$ dimensional crossovers 
[2-5,10-14].
A possibility of the RS phase to exist was experimentally studied in 
Q1D layered organic superconductors 
(TMTSF)$_2$X (X = PF$_6$ and X = ClO$_4$) by Naughton's and 
Chakin's groups [15-17].
Their experiments gave hints on a possibility for superconductivity 
to exceed significantly the quasi-classical upper critical field 
$H^{\parallel}_{c2}(0)$ - the effect predicted in Refs.[2-5,10-14] - 
but they were not able to confirm the appearance of the RS phase with 
$d T_c / d H > 0$.
Analogous experiments, performed on a Q2D superconductor
Sr$_2$RuO$_4$ [18], did not detect any stabilization of 
superconductivity at 
$H > H^{\parallel}_{c2}(0)$.

The main obvious difficulty in the above mentioned efforts to discover 
the RS phase is the Pauli spin-splitting destructive mechanism against 
superconductivity and the related Clogston paramagnetic
limiting field, $H_p$ [1].
It is absent only for some triplet superconducting phases which are
believed to exist in (TMTSF)$_2$X [2,15,4] and Sr$_2$RuO$_4$ [19]
superconductors. 
On the other hand, recently there have appeared the NMR measurements 
[20] in favor of a singlet nature of superconductivity in (TMTSF)$_2$ClO$_4$
material as well as some doubts [21] in a triplet nature of superconductivity
in Sr$_2$RuO$_4$ one.

The goal of our Letter is a three-fold one.
Firstly, we show that, although in layered paramagnetically limited
(singlet) superconductors the RS phase may not be characterized by 
$d T_c / d H > 0$ feature [2-4, 10-14], nevertheless the RS phase reveals 
itself as another unique phenomenon - paramagnetic 
intrinsic Meissner effect (PIME).
Secondly, we extend microscopical theory [3] to describe the most
important from experimental point of view $d$- and $s$-wave Q2D 
superconductors with $\xi_{\perp} < d$, where $\xi_{\perp}$ is a coherence 
length perpendicular to the conducting layers and $d$ is an inter-layer 
distance.
And thirdly, we suggest simple experimental methods to detect
the PIME phenomenon in Q2D organic and high-T$_c$ superconductors 
by using in-plane torque, the upper critical field, and magnetization 
measurements.
In particular, we demonstrate that in-plane anisotropy due to anisotropic
Ginzburg-Landau coherence lengths, which disappears in an intermediate 
region of magnetic fields (where the Lawrence-Doniach model is applicable), 
appears again in high magnetic fields as a consequence of the PIME phenomenon 
(see Figs. 1,2).
We suggest to measure in-plane anisotropy of the upper critical
field and magnetization as well as in-plane torque in high magnetic fields 
to discover the PIME and RS
phenomena.
For these purposes, we derive a free energy of a Q2D
superconductor with $\xi_{\perp} < d$ in a parallel magnetic field 
from the Gor'kov formulation [22-24] of the microscopic superconductivity 
theory.
Our results coincide with that of the Lawrence-Doniach model 
[25,26] only at low enough magnetic fields, $H \ll H^*$, where the Meissner 
effect is diamagnetic.
We show that, at high magnetic fields,  $H \sim H^* \leq H_p$, the field 
starts to shrink the Cooper pair "sizes" in perpendicular to conducting layers 
direction due to $3D \rightarrow 2D$ crossovers in a parallel magnetic
field.
The above mentioned $3D \rightarrow 2D$ crossovers of the Cooper 
pairs are not taken into account in the Lawrence-Doniach model and, 
as shown below, are responsible for the unique PIME 
phenomenon. 

Let us consider a layered superconductor with a Q2D electron spectrum,
\begin{equation}
\epsilon({\bf p})= \epsilon_{\parallel} (p_x, p_y) + 2 t_{\perp} \cos(p_z d) \ , 
\ \ \ t_{\perp} \ll \epsilon_F \ ,
\end{equation}
in a parallel magnetic field,
\begin{equation}
{\bf H} = (0,H,0) \  ,  \ \ \ \ {\bf A} = (0,0,-Hx) \ ,
\end{equation}
where $\epsilon_{\parallel} (p_x,p_y) \sim \epsilon_F$ is in-plane electron 
energy, $t_{\perp}$ is an overlapping integral of electron wave functions in 
a perpendicular to the conducting planes direction, $\epsilon_F$ is the
Fermi energy. 
Electron spectrum (1) can be linearized near 2D Fermi surface (FS),
$\epsilon_{\parallel} (p_x,p_y)=\epsilon_F$, in the following
way,
\begin{equation}
\epsilon({\bf p}) - \epsilon_F =  v_x(p_y) [p_x - p_x(p_y)] + 2 t_{\perp} \cos(p_z d) \ , 
\end{equation}
where $v_x(p_y) = \partial \epsilon_{\parallel} (p_x,p_y) / \partial p_y$ is
a velocity component and $p_x(p_y)$ is the Fermi momentum 
along $x$-axis.

In the gauge (2), electron Hamiltonian in a magnetic field can be
obtained from Eq.(3) by means of the Peierls substitution method, 
$p_x \rightarrow - i (d/dx)$, $p_z \rightarrow p_z + (e/c) H x$
[6].
Therefore, electron Green functions in a magnetic field  
satisfy the following differential equation,
\begin{eqnarray}
\biggl\{ i \omega_n - v_x (p_y) \biggl[ - i \frac{d}{dx} - p_x(p_y) \biggl]
+2 t_{\perp} \cos \biggl( p_z d + \frac{eHdx}{c} \biggl) + 2 \mu_B H s \biggl\}
\nonumber\\
\times G_{i \omega_n} (x, x_1;p_y,p_z;s) = \delta (x-x_1) \ ,
\end{eqnarray}
where $ \omega_n$ is the Matsubara frequency [22], $\mu_B$ is the Bohr
magneton, $s = \pm \frac{1}{2}$ is an electron spin projection along
quantization $y$-axis;
$\hbar \equiv 1$. 
It is important that Eq.(4) can be solved analytically.
As a result, we obtain,
\begin{eqnarray}
G_{i \omega_n} (x, x_1;p_y,p_z;s) = - i \frac{ sgn \  \omega_n}{v_x(p_y)}
 \exp \biggl[ - \frac{\omega_n (x-x_1)}{v_x(p_y)} \biggl] \exp[i p_x(p_y)(x-x_1)]
\nonumber\\
\times \exp \biggl[ \frac{2 i \mu_B s H (x-x_1)}{v_x(p_y)} \biggl]
\exp \biggl\{ \frac{ i \lambda(p_y)}{2} \biggl[ \sin \biggl( p_zd + \frac{eHdx}{c} \biggl) 
- \sin \biggl( p_zd + \frac{eHdx_1}{c} \biggl) 
\biggl] \biggl\} \ ,
\end{eqnarray}
where $\lambda (p_y) = 4 t_{\perp} c / e v_x (p_y) H d$.

Linearized gap equation determining superconducting transition 
temperature, $T_c(H),$ can be derived using Gor'kov equations 
for non-uniform superconductivity
[3,23,24]. 
As a result, we obtain,
\begin{eqnarray}
\Delta(x) = V \oint  \frac{d l}{v_{\perp}(l)} \ 
&&\int^{\infty}_{ |x-x_1| > |v_x(l)|/ \Omega} dx_1 \ \frac{2 \pi T }{v_x(l)
\sinh \biggl[ \frac{ 2 \pi T |x-x_1|}{ v_x (l)} \biggl] }
\cos \biggl[ \frac{2  \mu_B H (x-x_1)}{v_x (l)} \biggl]
\nonumber 
\end{eqnarray}
\begin{eqnarray}
&&\times J_0 \biggl\{ 2 \lambda (l)  \sin \biggl[ \frac{ e H d (x-x_1)}{2 c} \biggl]
\sin \biggl[ \frac{ e H d (x+x_1)}{2 c} \biggl] \biggl\} \  \ \Delta(x_1) \ ,
\end{eqnarray}
where integration in Eq.(6) is made along 2D contour, $\epsilon_{\parallel} (p_x,p_y)=\epsilon_F$, $v_{\perp}(l)$ is a velocity component perpendicular
to the 2D FS, $V$ is an effective electron-electron interactions 
constant, $\Omega$ is a cut-off energy.
[Note that, although Eq.(6) is derived for singlet $s$-wave superconductors,
it is also valid for $d$-wave superconductors [27] if we redefine properly
anisotropic coherence lengths and the effective interactions constant
$V$.]

We point out that Eq.(6) is the most general one among the existing 
equations to determine the parallel upper critical field in a layered
superconductor.
In particular, it takes into account the Bragg reflections and related
$3D \rightarrow 2D$ dimensional crossovers of electrons,
which move in the extended Brillouin zone in a parallel magnetic
field.
As shown in Ref. [7], the above mentioned quantum effects result 
in a momentum quantization law for an electron momentum component
along $x$-axis.
This a reason why the kernel of the integral Eq.(6) is periodic [2,4] with
respect to variables $x$ and $x_1$.
In the case, where the destructive Pauli spin-splitting effects against 
superconductivity are absent [i.e., at $\mu_B=0$ in Eq.(6)], Eq.(6) 
possesses a periodic solution for $\Delta(x)$ at any 
magnetic field.
In this case, superconductivity is stable in an arbitrary strong magnetic
field and exists at high fields in a form of the RS phase with
$d T_c / dH >0$.
In the case of a singlet superconductivity, which is considered in the 
Letter, the Pauli spin-splitting effects may eliminate the superconductivity
with $d T_c / dH >0$.
Nevertheles, in the latter case, the RS phase reveals itself as unusual 
anisotropy of the upper critical field and magnetization in high magnetic 
fields, $H > H^*  \sim (t_{\perp}/T_c)^{1/2} \phi_0 / \xi_x d 
\ll H_p$ 
(see Figs. 1,2).

As the most general equation, Eq.(6) contains Ginzburg-Landau and
Lawrence-Doniach descriptions as its limiting cases at low enough  
magnetic fields, $H \ll H^*$. 
For the so-called Josephson coupled layered superconductors with 
$\xi_{\perp} < d$ [25,26], Eq.(6) may be simplified and rewritten in the following differential
form,

\begin{equation}
\biggl[ \frac{T_c-T}{T_c} - 2.1 \biggl( \frac{\mu_B H}{\pi T_c} \biggl)^2
+ \xi^2_x \frac{d^2}{dx^2} - A(H)
+ B(H) \cos \biggl( \frac{2 x \omega_c}{v_F} \biggl)  \biggl]
\Delta(x) = 0 \ 
\end{equation}
with
\begin{equation}
A(H)  = \frac{8 t^2_{\perp}}{\omega^2_c} \biggl <  \biggl[ \frac{v_F}{v_x(l)} \biggl]^2 
\int^{\infty}_0  \frac{d z}{\sinh(z)} \sin^2 \biggl[ \frac{\omega_c}{4 \pi T_c}
\frac{v_x(l)}{v_F} z \biggl]  
\biggl>
\ 
\end{equation}
and
\begin{equation}
B(H) = \frac{8 t^2_{\perp}}{\omega^2_c} \biggl <  \biggl[ \frac{v_F}{v_x(l)} \biggl]^2 
\int^{\infty}_0  \frac{d z}{\sinh(z)} \sin^2 \biggl[ \frac{\omega_c}{4 \pi T_c}
\frac{v_x(l)}{v_F} z \biggl] 
\cos \biggl[ \frac{\omega_c}{4 \pi T_c} \frac{v_x(l)}{v_F} z \biggl]
\biggl> \ ,
\end{equation}
where 
\begin{equation}
\biggl <  ... \biggl> = \oint \frac{dl}{v_{\perp} (l)} \biggl(...\biggl) \biggl/ \oint \frac{dl}{v_{\perp}(l)} \ .
\end{equation}
[Here, $\omega_c = eHv_Fd/c$ is a characteristic frequency of an electron
motion along open FS (1) [3], $\xi_x = \sqrt{7 \zeta (3)} \bigl< v^2_x(l) \bigl>^{1/2} / 4 \pi T_c$
is in-plane Ginzburg-Landau coherence length, 
$\mu_BH \simeq \omega_c(H)  \ll \pi T_c$.]

Note that Eqs.(7)-(10) extend the Lawrence-Doniach model [25,26] to the
case of strong magnetic fields and can be called extended Lawrence-Doniach
equations.
In contrast to the traditional Lawrence-Doniach equations, the coefficients $A(H)$
and $B(H)$ in Eqs.(7)-(10) depend on a magnetic field, which means that a 
probability for the Cooper pair to jump from one conducting layer to another
depends on the field.
This important feature of Eqs.(7)-(10) is a consequence of shrinking 
of the Cooper pairs "sizes" due to $3D \rightarrow 2D$ dimensional
crossover in a parallel magnetic field
[2,3,7].

Below, we are interested in descriptions of the RS and PIME phenomena,
therefore, we consider Eqs.(7)-(10) at high magnetic fields.
It is possible to show that at $H \geq H^*$ the solution of Eq.(7) can
be represented as $\Delta(x) = \Delta = const $, which corresponds to the 
RS phase [2,3].
In this case, the corresponding second order term of a free energy with 
respect to the order parameter $\Delta$  can be written in the following
simple form,
\begin{equation}
F ^2(T,H) = - N(\epsilon_F) \biggl[ \frac{T_c(H)-T}{T_c} \biggl] \Delta^2 \ ,
\end{equation}
where
\begin{equation}
T_c(H) = T_c - 2.1  \frac{(\mu_B H)^2}{\pi^2 T_c}  - 2.1 \frac{t^2_{\perp}}{\pi^2T_c} 
+ 0.95 \frac{t^2_{\perp}}{\pi^2 T_c} \biggl( \frac{eHd}{c} \biggl)^2 \xi^2_x \ ,
\end{equation}
where $N(\epsilon_F)$ is a density of states per one electron spin 
projection at 
$\epsilon = \epsilon_F$. 

Note that the the first term in Eq.(12) describes destruction of 
a singlet superconductivity by the Pauli spin-splitting effects,
whereas the last term in Eq.(12) is responsible for the restoration of 
superconductivity at high magnetic fields and for the appearance of the RS 
phase and PIME 
phenomenon.
If we take into account that the fourth order term of a free energy with 
respect to the order parameter $\Delta$ can be calculated at $H \geq H^*$ 
in a standard manner [23], 
$F^4 = 7 \zeta (3) N(\epsilon_F) \Delta^4 / 16 \pi^2 T^2_c$,
then we can minimize the total free energy and find that
\begin{equation}
F (T,H) = - \frac{4 \pi^2}{7 \zeta(3)} N(\epsilon_F) [ T_c(H)-T]^2 \ .
\end{equation}

Magnetization can be found by a differentiation of the free energy (13)
with respect to a magnetic field,
\begin{equation}
M(T, H) = - \frac{\partial F(T,H)}{\partial H} = 
\frac{8}{7 \zeta (3)} N(\epsilon_F) \biggl( \frac{T_c-T}{T_c} \biggl)
\biggl[ -4.2 \ \mu^2_B + 1.9 \biggl( \frac{e t_{\perp} d \ \xi_x}{c} \biggl)^2 
\biggl] H\ ,
\end{equation}
Eqs.(12)-(14) are the main results of the Letter.
Note that, in Eq.(14), the first term corresponds to a destruction of superconductivity
 due to the Pauli spin-splitting effects, whereas the second term represents unusual paramagnetic orbital contribution to a magnetic moment (i.e., the 
PIME phenomenon).
It is important that $\xi_x$ in Eqs.(12)-(14) is anisotropic and depends on a direction of a magnetic field, since it is in-plane component of a coherence length 
perpendicular to the 
field.
Therefore, the RS and PIME effects in Eqs.(12)-(14) can be detected by 
measuring a torque provided that spin-splitting effects are isotropic
(see Eqs.1,2).

In conclusion, we discuss possible experiments to discover the 
PIME and RS phenomena.
The most direct method is to create such layered superconducting 
super-lattice, where $\omega_c(H) \gg \mu_B H$
[28].
The latter condition means that the orbital effects are more important 
than the Pauli spin-splitting ones.
Therefore, in this case, the increase of transition temperature (12)
and the paramagnetic Meissner effect (14) can be directly 
observed.
Nevertheless, in most real physical compounds with $\xi_{\perp} < d$,
$\omega_c(H) \simeq \mu_B H$ and, thus, the PIME (14) and RS (12)
phenomena can be observed only indirectly - by measurements of
anisotropies of the in-plane upper critical field (12) and magnetization (14) 
as well as by measurements of in-plane torque.
In our opinion, the most perspective superconductors for indirect 
observations  of the PIME phenomenon in steady magnetic fields are 
organic compounds $\alpha$-(ET)$_2$NH$_4$Hg(SCN)$_4$, 
$\kappa$-(ET)$_2$Cu(NCS)$_2$, 
$\kappa$-(ET)$_2$Cu[N(CN)$_2$]X, 
$\alpha$-(ET)$_2$KHg(SCN)$_4$, 
and $\lambda$-(BETS)$_2$FeCl$_4$
[29].
The above mentioned studies of the in-plane anisotropies can be performed 
also in high-temperature  superconductor Y$_1$Ba$_2$C$_3$O$_7$ but it 
will require ultra-high pulsed magnetic 
fields.

The author is thankful to N.N. Bagmet, P.M. Chaikin, and M.V. Kartsovnik 
for the 
useful discussions.
  
$^*$Also, Landau Institute for Theoretical Physics, 
2 Kosygina Street, Moscow, Russia.

\pagebreak

\begin{figure}[h]
\includegraphics[width=6.6in,clip]{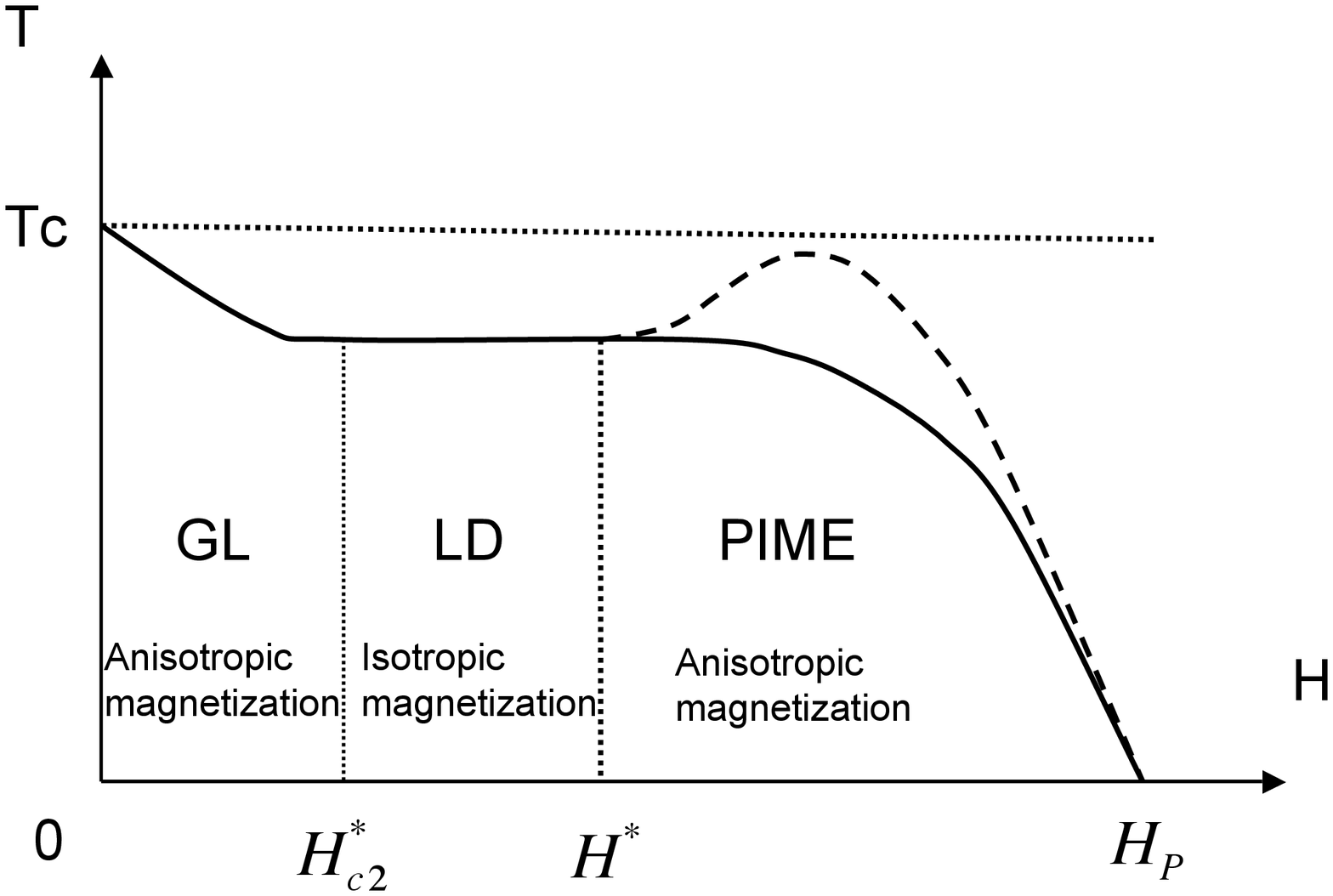}
\caption{Superconducting transition temperature in a parallel 
magnetic field for a paramagnetically limited Q2D superconductor 
is sketched.
GL - area of applicability of the Ginzburg-Landau theory [1],
LD - area of applicability of the Lawrence-Doniach model [25,26],
PIME - area, where both the GL and LD descriptions 
are broken.
In the latter case, which corresponds to shrinking of the Cooper pairs "sizes"
by a magnetic field, our Eqs. (6)-(14) are still valid and the Reentrant Superconducting 
(RS) phase appears.
The RS phase may reveal itself as an increase of the transition temperature 
in a magnetic field, if the orbital effects of an electron motion are stronger than 
the Pauli spin-splitting effects (dashed line). 
The RS phase always reveals itself as a paramagnetic intrinsic Meissner
effect (PIME), which results in unexpected in-plane anisotropy 
of the upper critical field and magnetization even in the case, where the 
Pauli spin-splitting effects are strong and, thus, the area with $d T_c/dH > 0$ 
is absent (solid line).
We suggest to measure in-plane torque, the upper critical field, and magnetization 
to discover the RS phase.}
\label{fig1}
\end{figure}

\begin{figure}[h]
\includegraphics[width=6in,clip]{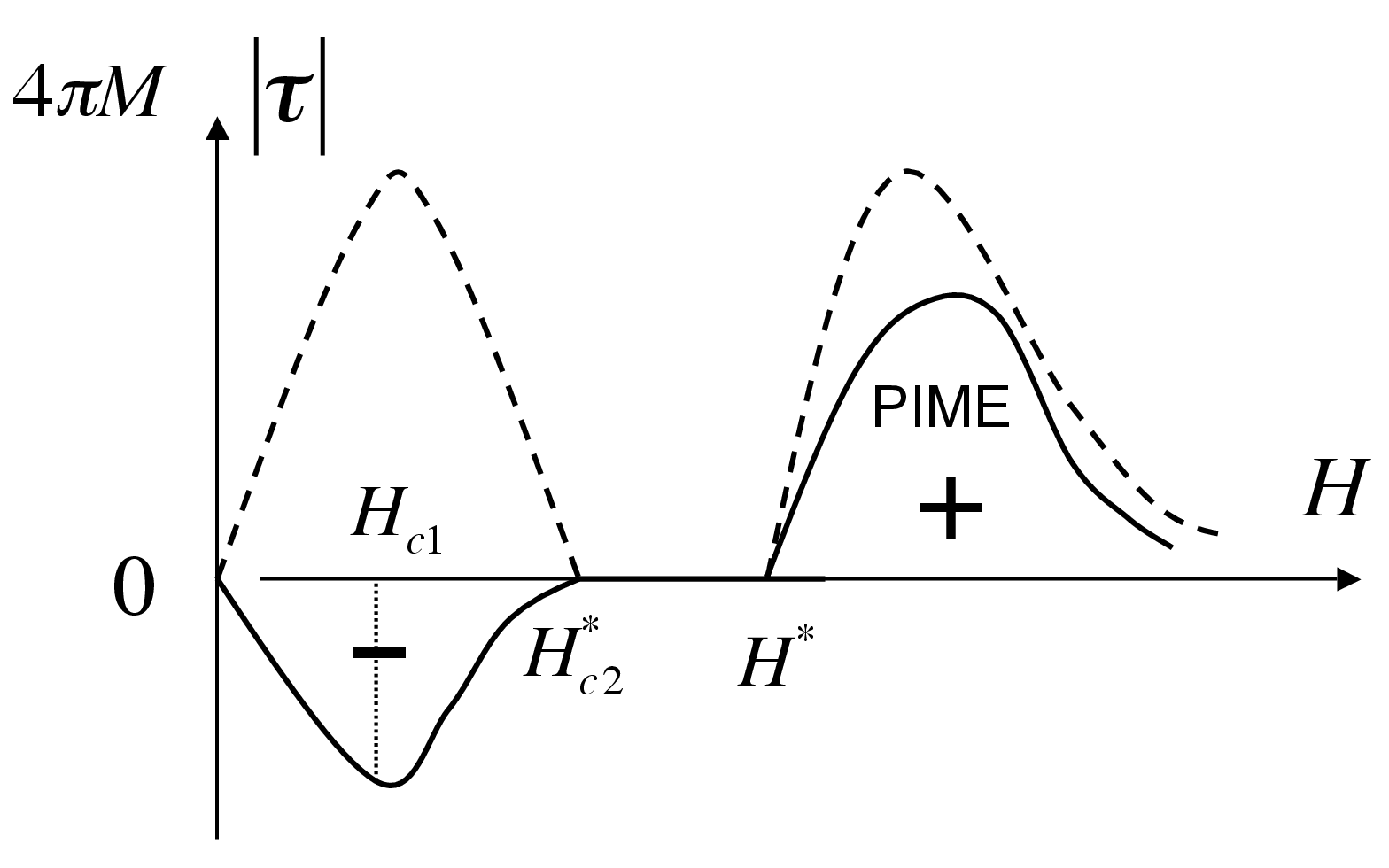}
\caption{Solid line: in-plane magnetization, $4 \pi {\bf M}$, is sketched
as a function of a magnetic field in the absence of the Pauli spin-splitting
effects.
At high magnetic fields, $H \geq H^*$, the Reentrant Superconducting (RS)
phase reveals itself as a paramagnetic intrinsic Meissner effect (PIME).
Dashed line: an absolute value of in-plane torque, $|{\bf \tau}|$, is 
sketched. 
It is important that the torque is independent on the Pauli spin-splitting
effects since they are isotropic. 
Therefore, even in the case, where the destructive Pauli spin-splitting effects 
eliminate a positive sign of the Meissner effect in high magnetic field, the PIME phenomenon and the RS phase can still be detected 
by the in-plane torque measurements.}  
\label{fig2}
\end{figure}


\begin{references}
\bibitem {Lebedd}
 In this Letter, we calculate the paramagnetic Meissner
moment and its two-fold anisotropy.
We disregard a weak four-fold anisotropy due to a $d$-wave
gap since it is $(t_{\perp} / T_c)^2 \ll 1$ times smaller than the
above calculated two-fold one.
\end{references}
\end{document}